\documentclass[8.5pt,twoside,twocolumn]{article}
\usepackage[super,sort&compress,comma]{natbib}
\usepackage{mhchem}
\usepackage{times,mathptmx}
\usepackage{sectsty}
\usepackage{balance}

\usepackage{graphicx} 
\usepackage{lastpage}
\usepackage[format=plain,justification=raggedright,singlelinecheck=false,font=small,labelfont=bf,labelsep=space]{caption}
\usepackage{fancyhdr}
\usepackage{color}
\usepackage{extarrows}

\newcommand{\ib}[1]{{\color{black}#1}}

\graphicspath{{ }{.}{./data/}}
\DeclareGraphicsExtensions{.eps}
\begin{document}

\thispagestyle{plain}
\fancypagestyle{plain}{
\renewcommand{\headrulewidth}{1pt}}
\renewcommand{\thefootnote}{\fnsymbol{footnote}}
\renewcommand\footnoterule{\vspace*{1pt}%
\hrule width 3.4in height 0.4pt \vspace*{5pt}}
\setcounter{secnumdepth}{5}

\makeatletter
\def\subsubsection{\@startsection{subsubsection}{3}{10pt}{-1.25ex plus -1ex minus -.1ex}{0ex plus 0ex}{\normalsize\bf}}
\def\paragraph{\@startsection{paragraph}{4}{10pt}{-1.25ex plus -1ex minus -.1ex}{0ex plus 0ex}{\normalsize\textit}}
\renewcommand\@biblabel[1]{#1}
\renewcommand\@makefntext[1]%
{\noindent\makebox[0pt][r]{\@thefnmark\,}#1}
\makeatother
\renewcommand{\figurename}{\small{Fig.}~}
\sectionfont{\large}
\subsectionfont{\normalsize}

\fancyfoot{}
\fancyfoot[RO]{\footnotesize{\sffamily{1--\pageref{LastPage} ~\textbar  \hspace{2pt}\thepage}}}
\fancyfoot[LE]{\footnotesize{\sffamily{\thepage~\textbar\hspace{3.45cm} 1--\pageref{LastPage}}}}
\fancyhead{}
\renewcommand{\headrulewidth}{1pt}
\renewcommand{\footrulewidth}{1pt}
\setlength{\arrayrulewidth}{1pt}
\setlength{\columnsep}{6.5mm}
\setlength\bibsep{1pt}

\newcommand{\alt}{\raisebox{-0.3ex}{$\stackrel{<}{\sim}$}}
\newcommand{\agt}{\raisebox{-0.3ex}{$\stackrel{>}{\sim}$}}

\twocolumn[
  \begin{@twocolumnfalse}
\noindent\LARGE{\textbf{Counterintuitive issues in the charge transport through molecular junctions}}
\vspace{0.6cm}

\noindent\large{\textbf{Ioan B\^aldea $^{\ast}$
\textit{$^{a\ddag}$}
}}\vspace{0.5cm}

\noindent \textbf{\small{Published: Phys.~Chem.~Chem.~Phys.~2015, {\bf 17}, 31260 - 31269 DOI: 10.1039/C5CP05476A}}
\vspace{0.6cm}

\noindent 
\normalsize{Abstract:\\
Whether at phenomenological or microscopic levels, most theoretical approaches to
charge transport through molecular junctions postulate or attempt to justify microscopically 
the existence of a dominant molecular orbital (MO). Within such single level descriptions,
experimental current-voltage $I-V$ curves are sometimes/often analyzed by using
analytical formulas expressing the current as
a cubic expansion in terms of the applied voltage $V$, and relate possible $V$-driven
shifts of the level energy offset relative to the metallic Fermi energy $\varepsilon_{0}$ 
to an asymmetry of molecule-electrode couplings 
or to an asymmetric location of the ``center of gravity'' of the MO with respect to electrodes.
In this paper, we present results demonstrating the failure of these intuitive expectations.
For example, we show how typical data processing based on cubic expansions 
yields a value of $\varepsilon_0$ underestimated
by a \ib{typical} factor of about two. \ib{When compared to theoretical results of DFT approaches, which typically 
underestimate the HOMO-LUMO gap by a similar factor, this may create the false impression of ``agreement'' with 
experiments in situations where this is actually not the case. Further,} such cubic expansions yield 
model parameter values dependent on the bias range width employed
for fitting, which is unacceptable physically. Finally, we present an example demonstrating that, counter-intuitively, 
\ib{the bias-induced change in the energy 
of an MO located much closer to an electrode can occur in a
direction that is opposite to the change in the Fermi energy 
of that electrode.
This is contrary to what one expects based on a ``lever rule'' argument, 
according to which the MO ``feels'' the local value of the electric potential, which is assumed 
to vary linearly across the junction and is closer to the potential of the closer electrode.
This example emphasizes the fact that screening effects in molecular junctions can have a 
subtle character, contradicting common intuition.}
$ $ \\  

{{\bf Keywords}: 
molecular electronics; molecular junctions; tunneling transport; Newns-Anderson model;
outer valence Green's function (OVGF) method; Stark effect
}
}
\vspace{0.5cm}
 \end{@twocolumnfalse}
  ]


\footnotetext{\textit{$^{a}$~Theoretische Chemie, Universit\"at Heidelberg, Im Neuenheimer Feld 229, D-69120 Heidelberg, Germany.}}
\footnotetext{\ddag~E-mail: ioan.baldea@pci.uni-heidelberg.de.
Also at National Institute for Lasers, Plasmas, and Radiation Physics, Institute of Space Sciences,
Bucharest-M\u{a}gurele, Romania}
%
%
\section{Introduction}
\label{sec:intro}
In spite of significant advances 
\cite{Tao:96,Reed:97,Nitzan:03,Venkataraman:06,Tao:06,Ratner:07a,Choi:08,Ruitenbeek:08,McCreery:09,Whitesides:10a,Reed:11,Wandlowski:11e,Metzger:15}, 
charge transport across molecular junctions continues to remain a nonequilibrium problem
difficult to understand \cite{Baldea:2014c,Baldea:2015c}. Resorting to 
a single (Newns-Anderson) model \cite{Anderson:61,Newns:69b,Schmickler:86,Datta:97,Metzger:01b,Datta:03,Datta:05,HaugJauho:08,Wandlowski:11e,Scheer:12,Baldea:2013b,Buimaga:13,Baldea:2015d} to describe the transport
within a picture assuming the existence of a dominant molecular orbital
is a common procedure in the field, also allowing to rationalize  
more sophisticated microscopic transport calculations \cite{CuevasScheer:10}. In fact, 
this single-level picture turned out to excellently explain 
a series of transport measurements beyond the ohmic bias range 
\cite{Baldea:2012a,Baldea:2012b,Baldea:2012g}
and to back the model parameters extracted from fitting experimental data with 
high-level quantum chemical calculations \cite{Baldea:2013b,Baldea:2014e,Baldea:2015d}.
In the present work, we consider two issues related to the analysis of the transport data
within this framework:

(i) Typical current-voltage $I-V$ characteristics measured in molecular junctions are 
featureless curves. An example is depicted in \figurename\ref{fig:iv-Tan} below.  
Due to their general appearance, although a general analytic formula $I=I(V)$ 
is available from literature 
\cite{Schmickler:86,HaugJauho:08,Metzger:01,CuevasScheer:10,Baldea:2010e}, 
using third-order expansions instead of the exact expression
(see eqn~(\ref{eq-j-arctan}) below) appears to be a convenient and reasonable 
simplification and was used in earlier studies \cite{Wandlowski:11e,Scheer:12}.

Such third-order expansions are inspired by studies on a variety of macroscopic and mesoscopic 
junctions up to biases of current experimental interest, 
wherein it was considered a prominent characteristic of transport via tunneling
\cite{Simmons:63,Simmons:63d,Rowell:69,Duke:69b,Duke:69,Brinkman:70}.

(ii) Like those shown in \figurename\ref{fig:iv-Tan}, experimental $I-V$ curves  
may exhibit a more or less pronounced asymmetry upon bias 
polarity reversal $I(-V) \neq - I(V)$, which is particularly desirable 
for achieving current rectification using molecular devices \cite{Metzger:15}. The most common way 
to embody this asymmetry in analytic transport approaches is 
either to relate it to asymmetric 
molecule-electrode couplings \cite{CuevasScheer:10,Scheer:12} or to assume that 
the (``center of gravity'' of the) dominant molecular orbital is located 
asymmetrically relative to the two (say, ``substrate'' s and ``tip'' t) 
electrodes \cite{Metzger:01,Datta:03}.

The analysis presented below will demonstrate that, although the aforementioned assumptions 
seem to be justified intuitively, they are in fact of rather limited applicability. Examples will
be presented showing cases where the opposite is true.
\section{Results and discussion}
\label{sec:results}
\ib{To provide the reader with a convenient reference when reading the text that follows, 
the definition of the variables utilized below is given in 
Table \ref{table}.
\begin{table}[h]
\small
\ib{
  \caption{\ib{List of the main variables utilized in the present paper.}}
  \label{table}
  \begin{tabular*}{0.5\textwidth}{@{\extracolsep{\fill}}ll}
    \hline
    Symbol & Meaning \\
    \hline
    MO & (dominant) molecular orbital\\
    $\varepsilon_{0}(V)$ & MO energy offset under applied bias ($V\neq 0$)\\
    $\varepsilon_{0} \equiv \left\vert\varepsilon_{0}(V)\right\vert_{V=0}$ & 
    MO energy offset without applied bias or for\\
      & junctions with symmetric $I-V$ curves\\
    $\gamma$ & (dimensionless) Stark effect strength\\
      & ($-1/2 \leq \gamma \leq 1/2$)\\
    $\Gamma_{s,t}$ & MO broadening functions due to coupling to\\
      & electrodes $s$ (``substrate'') and $t$ (``tip'') \\
    $\Gamma_{a}$ & arithmetic average of $\Gamma_{s}$ and $\Gamma_{t}$ \\ 
    $\Gamma_{g}$ & geometric average of $\Gamma_{s}$ and $\Gamma_{t}$ \\ 
    $\Gamma_{h}$ & harmonic average of $\Gamma_{s}$ and $\Gamma_{t}$ \\ 
    $V_t\left(\equiv V_p\right)$ & transition voltage (alias peak voltage, 
                                 \emph{cf.}~ref.~\citenum{Baldea:2015b}\\
          & and \citenum{Baldea:2015c}) for junctions with symmetric $I-V$ curves \\
    $I_t \equiv I(V_t)$ & current at $V=V_t$ \\
    $V_{t,\pm}\left(\equiv V_{p,\pm}\right)$ & transition (peak) voltages for positive/negative biases\\
    $V_{t}^{j} \ (j=1,2)$ &  transition voltage for positive/negative biases;\\
                         & $V_{t}^{1,2} = V_{t,\pm} \mbox{\,sign\,} \varepsilon_0$\\
    $N$ & number of molecules in junction \\
    $I_{3}$ & current within the cubic expansion (approx.)\\
    $V_{t,3}^{j} \ (j=1,2)$ &  transition voltage for positive or negative bias\\ 
            & within the cubic approximation\\
    $\delta$ & asymmetry of the MO-electrode couplings\\
             & ($0 < \delta < 1$)\\
    $G$ & low bias conductance \\
    $G_0$ & conductance quantum \\
    $\mu_{s,t}$ & electrodes' Fermi energy \\
    $G^{fit}, \varepsilon_{0}^{fit}$ & conductance and MO energy offset deduced by\\
        & fitting using cubic expansions $I\ vs.\ V$\\
    $V_b$ & bias range used for fitting\\ 
    $\varepsilon_{0}^{fit}\left(V_b\right)$ & MO energy offset deduced by fitting $I-V$ curves\\
        & obtained experimentally of \emph{via} eqn (\ref{eq-i-ib}) using cubic\\
        & expansions in the bias range $-V_b < V < V_b$\\
    $\delta \varepsilon_{0}$ &  bias-driven MO shift\\
    $z$ & molecular axis\\
    $d$ & molecular length\\
    $z_0$ & MO ``center of gravity'' location\\
    $E_z$ & electric field along the molecular axis\\
    $\delta \varepsilon_{0}(z)$ & bias-driven MO shift expected according to the\\
        & ``lever rule'' (\emph{cf.}~ref.~\citenum{Metzger:01b})\\
    \hline
  \end{tabular*}
}
\end{table}
}

\subsection{Basic working equations}
\label{sec:equations}
By assuming electrode bandwidths much larger than all other characteristic 
energies (wide band limit), 
the current mediated by a single, possibly bias-dependent energy 
level having an energy offset $\varepsilon_0(V)$ relative to the 
electrodes equilibrium Fermi energy can be written in a compact form 
\begin{equation}
\label{eq-j-1-arctan}
I = N G_0 \frac{\Gamma_h}{e} 
\arctan\frac{e V / \Gamma_a}{1 + \frac{\left[\varepsilon_0(V)\right]^2 - e^2 V^2/4}{\Gamma_a^2}}
\end{equation}
The above formula results by recasting the more familiar expression 
\cite{Schmickler:86,HaugJauho:08,Metzger:01,CuevasScheer:10,Baldea:2010e}
\begin{equation}
\label{eq-j-arctan}
I =  N G_0 \frac{\Gamma_h}{e}
\left[
\arctan\frac{\varepsilon_0(V) + e V/2}{\Gamma_a}
-
\arctan\frac{\varepsilon_0(V) - e V/2}{\Gamma_a}
\right]
\end{equation}
with the aid of the trigonometric identity
\begin{equation*}
\arctan a - \arctan b = \frac{a - b}{1 + a b}
\end{equation*}
Above, the biased (``substrate'' $s$ and ``tip'' $t$) electrodes 
are assumed to have Fermi energies $\mu_{s,t} = \pm eV/2$, 
$N$ is the effective number of molecules in junction, and 
$G_0 = 2 e^2/h$ is the conductance quantum. $\Gamma_{h,a,g}$ stand for 
the harmonic, arithmetic, and geometric averages of the level broadening 
functions $\Gamma_{s,t}$ arising from the couplings between molecule and 
electrodes. 
In the zero-bias limit ($V \to 0$), the conductance $G$ has the form
\begin{equation}
G = N G_0 \frac{\Gamma_g^2}{\varepsilon_{0}^2 + \Gamma_a^2} 
\label{eq-G}
\end{equation}
By assuming a bias-independent level energy offset, the third-order expansion 
$\mathcal{O}\left( V\right)^3$ 
in terms of $V$ of eqn (\ref{eq-j-arctan}) or (\ref{eq-j-1-arctan}) reads
\begin{equation}
I \approx I_{3} = G V
\left[
1 + 
\frac{(e V)^2 \left(3 \varepsilon_0^2 -\Gamma_{a}^2\right)}{12 \left(\varepsilon_0^2+\Gamma_{a}^2\right)^2}
\right]
\label{eq-i3-sym}
\end{equation}
In the off-resonance limit
($\Gamma_a \ll \vert \varepsilon_0\vert$) 
which characterizes the vast majority of experimental situations, the above expression
acquires the form 
\begin{equation}
I \approx I_{3} = G V \left[ 1 +
     \left(\frac{e V}{2 \varepsilon_{0}}\right)^2 
\right] 
\label{eq-i3-sym-off}
\end{equation}

An applied bias $V$ can shift the energy of the dominant orbital,
$\varepsilon_0 \to \varepsilon_0(V)$. 
By assuming a linear dependence
\begin{equation}
\varepsilon_0(V) = \varepsilon_0 + \gamma e V 
\label{eq-gamma}
\end{equation}
a series of experiments could be successfully analyzed.
In this case, the counterpart of the third-order expansions of eqn (\ref{eq-i3-sym}) 
and (\ref{eq-i3-sym-off}) read
\begin{equation}
I \approx I_{3} = G V
\left[1 - 
2  \gamma \frac{\varepsilon_0  e V}{\varepsilon_0^2 + \Gamma_a^2} +
\left(\frac{1}{12} + \gamma^2\right)\frac{3 \varepsilon_0^2 - \Gamma_a^2}
{\left(\varepsilon_0^2 + \Gamma_a^2\right)^2} ( e V)^2
\right]
\label{eq-i3-asym}
\end{equation}
and
\begin{equation}
I \approx I_{3} = G V \left[ 1 -2 \gamma \frac{eV }{\varepsilon_{0}}+
     \frac{1 + 12 \gamma ^2}{4}
     \left(\frac{e V}{\varepsilon_{0}}\right)^2 
\right]
\label{eq-i3-asym-off}
\end{equation}
respectively. 

In off-resonance cases 
($\Gamma_a \ll \vert \varepsilon_0 \pm eV/2 \vert $), an expression  
for the current \emph{not} limited to low-order expansions in $V$ 
can be deduced from eqn (\ref{eq-j-arctan}) \cite{Baldea:2012a}
\begin{equation}
I = G V \frac{\varepsilon_0^2}{ \left[\varepsilon_0(V)\right]^2 -  e^2 V^2/4}
\label{eq-i-ib}  
\end{equation}

To end this section, we briefly refer to a quantity useful for the subsequent analysis,
namely the transition voltage $V_t$, defined as the bias at the minimum of the Fowler-Nordheim quantity
$\log \left(\vert I\vert/V^2\right)$, or the equivalent peak voltage $V_p (\equiv V_t)$, 
defined as the bias
at the maximum of $V^2/\vert I\vert$. The latter has been recently introduced 
\cite{Baldea:2015b,Baldea:2015c} to emphasize that no mechanistic transition (\emph{e.g.},
from direct tunneling to field-emission tunneling, as initially claimed \cite{Beebe:06})
occurs at $V=V_t \equiv V_p$. 

In off-resonant situations described by eqn (\ref{eq-i-ib}) and (\ref{eq-i3-sym-off}) 
or (\ref{eq-i3-asym-off}),
simple expressions for the transition (peak) voltages for both polarities 
($V_{t}^{1} V_{t}^{2} < 0$) 
can be derived analytically \cite{Baldea:2012a,Baldea:2015c}
\begin{equation}
\mbox{Eqn (\ref{eq-i-ib})} \Rightarrow 
e V_{t}^{1,2} = \pm \frac{\varepsilon_0}{\sqrt{\gamma^2 + 3/4} - 2 \gamma} 
\xlongrightarrow[\text{}]{\text{$\gamma \to 0$}} 
e V_{t} \equiv \vert V_{t}^{1,2}\vert = \frac{2}{\sqrt{3}} \vert \varepsilon_0 \vert
\label{eq-vt}  
\end{equation}
\begin{equation}
\mbox{Eqn (\ref{eq-i3-sym-off}) or (\ref{eq-i3-asym-off})} \Rightarrow 
e V_{t,3}^{1,2} = \pm \frac{2}{1 + 12 \gamma^2} \varepsilon_0
\xlongrightarrow[\text{}]{\text{$\gamma \to 0$}}  
e V_{t,3} \equiv \vert V_{t,3}^{1,2}\vert = 2 \vert \varepsilon_0 \vert
\label{eq-vt3}  
\end{equation}
We checked that the off-resonance limit applies in all the cases presented below.
Therefore, using the simplified eqn (\ref{eq-i-ib}) instead of eqn (\ref{eq-j-arctan}) 
or (\ref{eq-j-1-arctan}), and eqn (\ref{eq-i3-sym-off}) or (\ref{eq-i3-asym-off}) instead of
eqn (\ref{eq-i3-sym}) or (\ref{eq-i3-asym}) is legitimate. 
\subsection{Exact Newns-Anderson description \emph{versus} cubic expansion}
\label{sec:cubic}
By fitting experimental $I-V$-curves using eqn (\ref{eq-i-ib}) and (\ref{eq-i3-asym-off}),
the values of the fitting parameters $\varepsilon_{0}$, $\gamma$, and $G$ entering these equations
can be deduced. The analysis presented in this subsection will reveal 
surprising differences between the values estimated with the aid of these equations.
   
\begin{figure}
\centerline{
\includegraphics[width=0.45\textwidth]{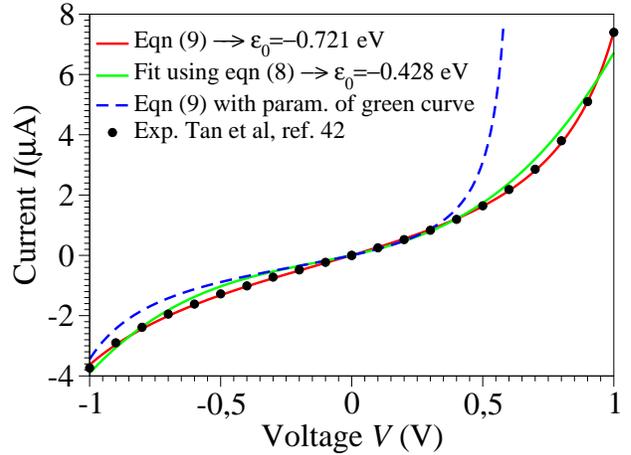}
}
\caption{
Raw experimental current-voltage data (courtesy of Pramod Reddy) \cite{Tan:10} fitted with eqn (\ref{eq-i-ib})
(red curve) and with the third-order approximation of eqn (\ref{eq-i3-asym-off}) (green curve). 
The parameter values are $\varepsilon_0 = -0.721$\,eV, $\gamma = 0.065$, and $G=2.575\,\mu$S
for the red curve and 
$\varepsilon_0 = -0.428$\,eV, $\gamma = 0.165$, and $G=1.922\,\mu$S
for the green curve. As visible in the figure, the blue dashed line, obtained by using eqn (\ref{eq-i-ib})
and the parameters corresponding to the green line, substantially deviates from the green curve, which would
not be the case if the cubic approximation of eqn (\ref{eq-i3-asym-off}) were justified.
The red curve could hardly be distinguished within the drawing
accuracy from that obtained using eqn (\ref{eq-j-1-arctan}) or (\ref{eq-j-arctan}), because
of the small values $\Gamma_s \approx \Gamma_t \sim 10^{-2}\vert \varepsilon_0\vert$ deduced from
eqn (\ref{eq-G}) with $N\approx 100$ \cite{Tan:10}. The negative $\varepsilon_0$-values
reflect the HOMO-mediated conduction in the junctions considered \cite{Tan:10}.
}
\label{fig:iv-Tan}
\end{figure}

The black symbols of \figurename\ref{fig:iv-Tan} depict a typical, moderately 
asymmetric ($I(-V) \neq - I(V)$) curve measured in molecular junctions \cite{Tan:10}.
Fitting the experimental data shown in \figurename\ref{fig:iv-Tan} with the aid 
of eqn (\ref{eq-i-ib}) yields a
curve (red line in \figurename\ref{fig:iv-Tan}) in virtually perfect agreement
with experiment. 
Although not so ``perfect'' as the red line, the green curve, 
obtained by fitting the experimental data using eqn (\ref{eq-i3-asym-off}), 
is in fact quite satisfactory. Still, much more importantly than the quality of the two fits,
the two fitting procedures yield substantially different
parameter values (see caption of \figurename\ref{fig:iv-Tan}). Particularly noteworthy
is the fact that the cubic approximation drastically underestimates the HOMO-energy
offset $\varepsilon_0$, which represents $\sim 60$\% from the exact estimate. 
(Let us mention that in the junctions considered conduction is mediated by the highest occupied 
molecular orbital (HOMO)\cite{Tan:10}). 

\ib{What is wrong with the 
green curve of \figurename\ref{fig:iv-Tan} is the fact that the cubic expansion
leading to eqn (\ref{eq-i3-asym-off}) is a legitimate approximation of eqn (\ref{eq-j-arctan}) 
or (\ref{eq-j-1-arctan}), or eqn (\ref{eq-i-ib}) (because in the present off-resonant limit 
($\Gamma_a \ll \vert \varepsilon_0\vert$, see caption of \figurename\ref{fig:iv-Tan}) 
eqn (\ref{eq-j-arctan}) or (\ref{eq-j-1-arctan}) reduce to eqn (\ref{eq-i-ib}))
\emph{only} at sufficiently low biases. If this were the case, the differences between 
the $I-V$ curves computed using 
eqn (\ref{eq-i3-asym-off}) (green curve in  \figurename\ref{fig:iv-Tan}) and eqn (\ref{eq-i-ib})
(blue curve in \figurename\ref{fig:iv-Tan}) at the \emph{same} parameter values would be negligible.
However, the inspection of \figurename\ref{fig:iv-Tan} clearly reveals that this is not the case. The cubic
expansion does not hold in the whole experimental $V$-range; 
it is legitimate only up to biases $\vert V\vert \sim 0.3$\,V,
wherein the differences between the green and blue curves are small.}

\begin{figure}
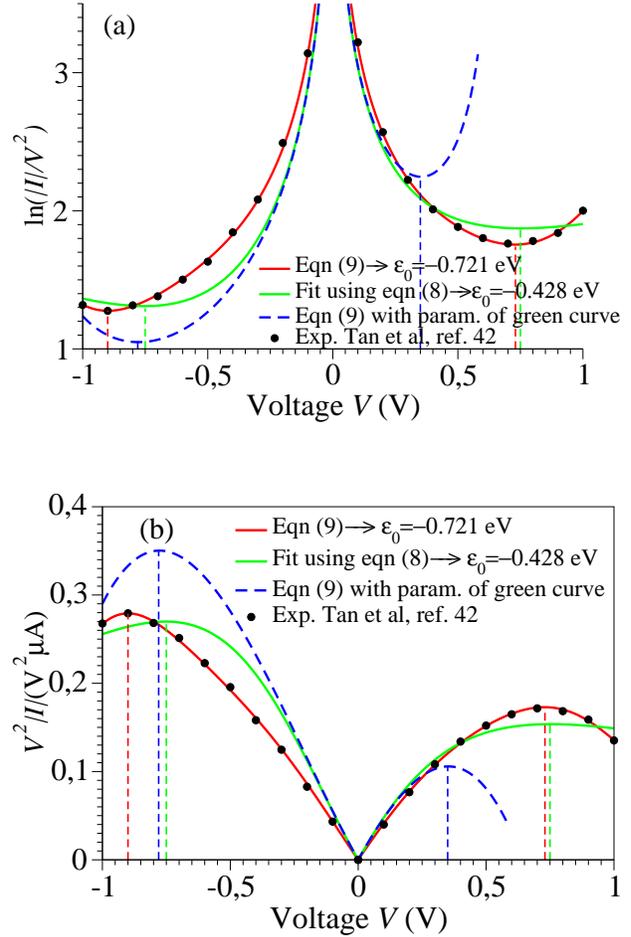

\centerline{\includegraphics[width=0.45\textwidth]{fig2a}}
$ $\\[3ex]
\centerline{\includegraphics[width=0.45\textwidth]{fig2b}}
\caption{The 
$I-V$ curves depicted in 
\figurename\ref{fig:iv-Tan} recast as transition (TVS, panel a) and peak (PVS, panel b) voltage spectra.
Notice the inability of the cubic approximation (green curves) to account for the experimental fact 
(\emph{cf.}~black symbols) that transition (peak) voltages ($V_{t,\pm} = V_{p,\pm}$) of opposite polarity 
--- which specify the location of the minima (maxima) in panel a (panel b) 
can have different magnitudes.
}
\label{fig:tvs-pvs-Tan}
\end{figure}

Using experimental transport data from ref.~\citenum{Tan:10}, \figurename\ref{fig:tvs-pvs-Tan} 
emphasizes another important drawback of the cubic approximation, namely its inability to account for
the \emph{experimental} fact that the transition (\figurename\ref{fig:tvs-pvs-Tan}a) or, 
alternatively, the peak (\figurename\ref{fig:tvs-pvs-Tan}b) voltage spectra have minima
or maxima located asymmetrically ($V_{t,+} \neq - V_{t,-}$). This result, based on experimental measurements,
gives further support to a similar finding emerging from 
a theoretical simulation presented recently \cite{Baldea:2015c}.

\begin{figure}
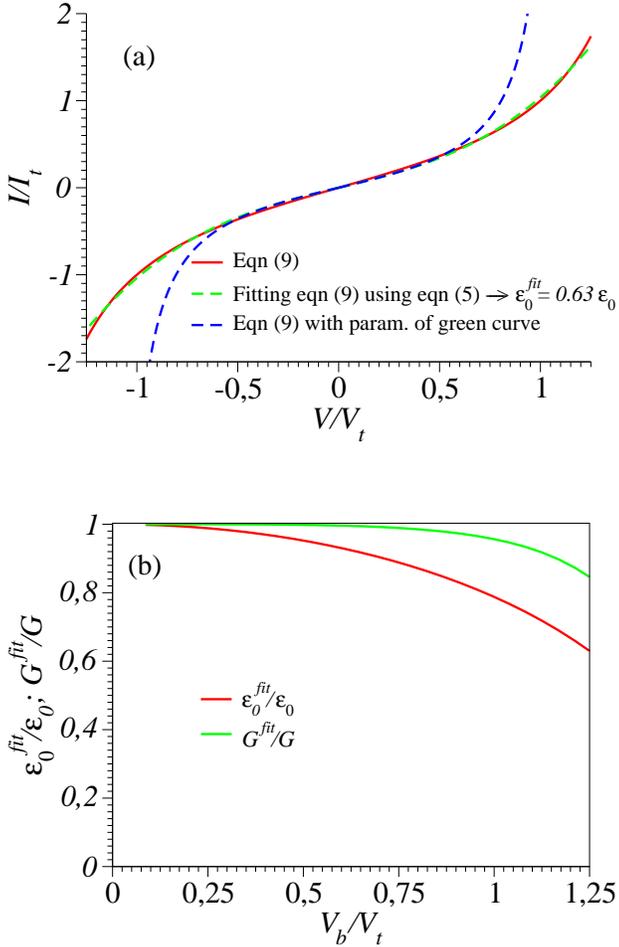

\centerline{\includegraphics[width=0.45\textwidth]{fig3a}}
$ $\\[3ex]
\centerline{\includegraphics[width=0.45\textwidth]{fig3b}}
\caption{(a)The current-voltage curve computed for $\gamma=0$ using eqn (\ref{eq-i-ib}) (red line)
and fitted (green line) with the aid of eqn (\ref{eq-i3-sym-off}) in the bias range shown
($0 < 1.25 V_t$). Notice that in spite of the very good quality of the fit,
this procedure substantially underestimates the $\varepsilon_0$-value
(given in the legend). (b) Values of $G^{fit}$ and $\varepsilon_{0}^{fit}$ 
obtained by fitting with the aid of eqn (\ref{eq-i3-sym-off}) 
the current computed using eqn (\ref{eq-i-ib}) in bias ranges $- V_b < V < V_b$. $V_b$ 
is the variable entering the abscissa. The reduced variables $I/I_t$  
and $V_b/V_t$ are expressed using the units 
$V_t = \left(2/\sqrt{3}\right) \varepsilon_0/e $ of eqn (\ref{eq-vt}) and 
$I_t = I\left(V_t\right) = G\vert \varepsilon_0\vert \sqrt{3}/e $ \cite{Baldea:2015b}.
}
\label{fig:iv-e0-gamma=0}
\end{figure}

Typical transport measurements on molecular junctions sample bias ranges slightly 
exceeding the transition voltage ($V \agt V_t$) \cite{Beebe:06,Beebe:08,Reed:09,Baldea:2010h}. 
In \figurename\ref{fig:iv-e0-gamma=0} we present results of a numerical simulation. 
There, the (red) curve has been computed using eqn (\ref{eq-i-ib})
at $\gamma=0$ (\emph{i.e.}, for a curve $I(-V)= -I(V)$ symmetric about origin $V=0$)
for such a bias range ($-1.25\,V_t < V < 1.25\,V_t $) 
along with the (green) curve obtained by fitting the red curve by means of eqn (\ref{eq-i3-sym-off}).
Similar to \figurename\ref{fig:iv-Tan}, the quality of the fit is very good;
nevertheless, the fitting 
parameter $\varepsilon_{0}^{fit}$ (63\% of the actual value $\varepsilon_0$) 
is drastically underestimated. For convenience, the results presented in 
\figurename\ref{fig:iv-e0-gamma=0} are presented in dimensionless variables obtained by using
the ``natural'' bias and current units $V_t$ and $I_t \equiv I\left(V_t\right)$
introduced recently \cite{Baldea:2015b}.    

An important pragmatic merit of the transition voltage is its reproducibility 
\cite{Frisbie:11,Guo:11,Reddy:11,Baldea:2012g}: in contrast to the very broad
conductance (or current \cite{Vuillaume:12a}) histograms, 
$V_t$-histograms are considerably narrower. 
Therefore, estimating the energy offset $\varepsilon_0$ from $V_t$ in cases where 
the existence of a single dominant level can be justified microscopically for the junction(s) 
in question (\emph{e.g.}, ref.~\citenum{Baldea:2013b}, \citenum{Baldea:2014e} and \citenum{Baldea:2015d})
may appear preferable to fitting $I-V$ data. 
From the experimental values $V_t = 1.15 \pm 0.15; 1.0 \pm 0.07; 0.87 \pm 0.07$\,V 
for molecular junctions of 
phenyldithiol and Ag; Au; Pt-electrodes, eqn (\ref{eq-vt}), which follows from 
eqn (\ref{eq-i-ib}) \cite{Baldea:2012a}, yields the (HO)MO energy offsets values 
$\vert\varepsilon_0 \vert = 1.0 \pm 0.1; 0.88 \pm 0.05; 0.75 \pm 0.04 $\,eV, respectively 
($\gamma=0$) \cite{Baldea:2015d}. These values agrees well with 
those deduced \emph{via} ultraviolet photoelectron spectroscopy (UPS):
$\vert\varepsilon_0 \vert = 1.1 \pm 0.1; 0.9 \pm 0.1; 0.8 \pm 0.1$\,eV, respectively \cite{Frisbie:11}.
By contrast, the values obtained \emph{via} eqn (\ref{eq-vt3}), which follows from eqn (\ref{eq-i3-sym-off}),
namely, $\vert\varepsilon_0 \vert = 0.57; 0.50; 0.43$\,eV, respectively 
are underestimated by a factor 
$\sim 58$\%. 
Noteworthy, the value of this factor is very close to those of the two aforementioned cases
(namely, $\sim 60\%$ and $\sim 63 \%$).

\ib{Like those presented in  \figurename\ref{fig:iv-Tan},} the results presented \ref{fig:iv-e0-gamma=0}a  
\ib{reemphasize} why attempting to fit transport data in bias ranges sampled in typical experiments 
by using the cubic approximation represents an inadequate procedure:
this $V$-range is \emph{beyond} the applicability of the cubic expansion. 
If the cubic expansion were legitimate, differences between curves computed via eqn 
(\ref{eq-i3-sym-off}) or (\ref{eq-i3-asym-off}) 
and the (practically) exact eqn (\ref{eq-i-ib})  
using the \emph{same} parameters deduced \emph{via} cubic fitting would be negligible.
(In \figurename\ref{fig:iv-Tan}
the differences between the red curve and experimental data \cite{Tan:10} are insignificant,
so here we could refer to the results computed \emph{via} eqn (\ref{eq-i-ib}) 
as the ``experimental'' results.) The comparison between 
solid green lines (cubic fitting) and the blue dashed lines 
(exact equation + parameters from cubic fitting) 
of \figurename\ref{fig:iv-Tan} and \ref{fig:iv-e0-gamma=0}a,
which depict the two aforementioned curves, shows that the opposite is true. 
These differences are small at low biases only; the contributions of the higher order terms 
neglected in eqn (\ref{eq-i3-sym-off}) and (\ref{eq-i3-asym-off})) 
is witnessed as significant differences between the green and blue curves at higher biases.

\begin{figure}
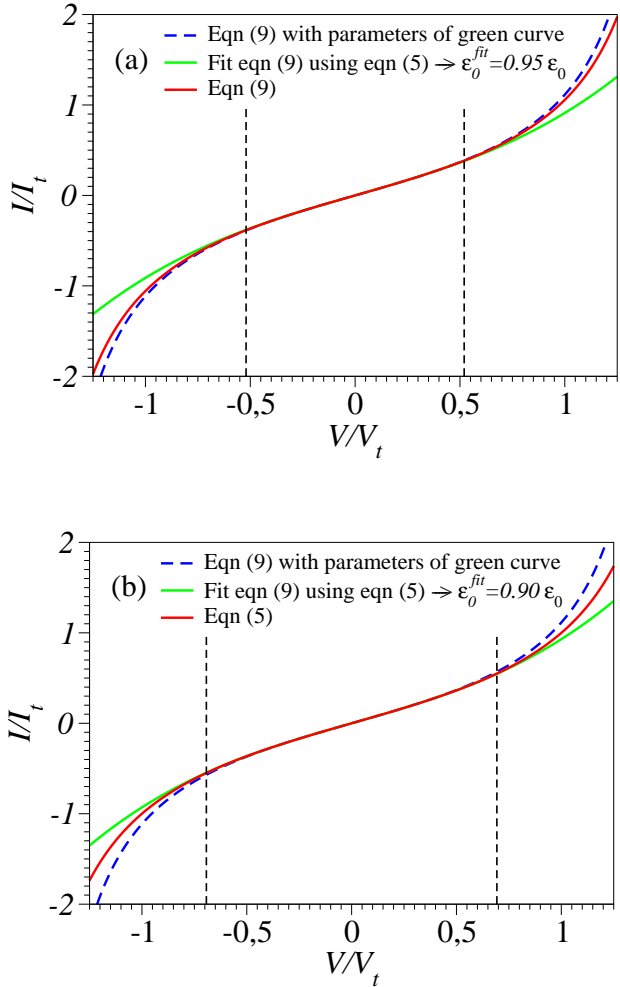

\centerline{
\includegraphics[width=0.45\textwidth]{fig4a}}
$ $\\[3ex]
\centerline{
\includegraphics[width=0.45\textwidth]{fig4b}
}
\caption{
The current-voltage curve computed for $\gamma=0$ using eqn (\ref{eq-i-ib}) (red line),
which mimics a symmetric ``experimental'' curve,
fitted with the aid of eqn (\ref{eq-i3-sym-off}) in the bias range delimited by the vertical dashed lines 
($-0.520 < V/V_t < 0.520$ in panel a and $-0.693 < V/V_t < 0.693$ in panel b). 
The reduced variables $I/I_t$ and $V/V_t$ are expressed using the units 
$V_t = \left(2/\sqrt{3}\right) \varepsilon_0/e $ of eqn (\ref{eq-vt}) and 
$I_t = I\left(V_t\right) = G\vert \varepsilon_0\vert \sqrt{3}/e $ \cite{Baldea:2015b}.
These results show that the level energy offset $\varepsilon_{0}^{fit}$ 
deduced \emph{via} fitting using cubic expansions restricted to sufficiently narrow bias ranges 
may represent acceptable estimates of the exact value $\varepsilon_{0}$. However, 
as explained in the main text, this procedure of restricting the bias range used for fitting 
cannot be applied for noisy curves.
}
\label{fig:iv-e0-gamma=0_Vmax}
\end{figure}

In this vein, one can attempt to employ cubic expansions 
for data fitting in narrower bias ranges, where higher order
terms are indeed negligible. Simulations of this kind are presented 
in \figurename\ref{fig:iv-e0-gamma=0}b, \ref{fig:iv-e0-gamma=0_Vmax}a and 
\ref{fig:iv-e0-gamma=0_Vmax}b. 
For simplicity, in these figures we have chosen
the case $\gamma=0$ ($I(-V)=-I(V)$), so fittings in a bias range $-V_b < V < V_b$
and $0 < V < V_b$ yield identical results. 
In \figurename\ref{fig:iv-e0-gamma=0_Vmax}a and 
\ref{fig:iv-e0-gamma=0_Vmax}b, we present results 
obtained from fitting by means of the cubic approximation of eqn (\ref{eq-i3-sym-off}) 
symmetric $I-V$ curves
(corresponding to $\gamma=0$) computed using the ``exact'' eqn (\ref{eq-i-ib})
(which mimic the ``experimental'' curves in these simulations)
within bias ranges $\vert V\vert < 0.6\,V_t$ and $\vert V\vert < 0.8\,V_t$, respectively.
Differences ($\sim 5\%$ and $10\%$, respectively) between the exact values of 
MO energy offsets ($\varepsilon_{0}$) 
and those ($\varepsilon_{0}^{fit}$) estimated in this way 
($\varepsilon_{0}^{fit} \simeq 0.95\,\varepsilon_{0}$ and 
$\varepsilon_{0}^{fit} \simeq 0.90\,\varepsilon_{0}$, respectively) are reasonable
and comparable to experimental inaccuracies (\emph{e.g.}, ref.~\citenum{Baldea:2015d}).
The examples depicted in \figurename\ref{fig:iv-e0-gamma=0_Vmax}a and 
\ref{fig:iv-e0-gamma=0_Vmax}b pass the
self-consistency test: the differences between the fitting curves 
(green curves, cubic expansions) and the
(blue) curves computed \emph{via} the exact eqn (\ref{eq-i-ib}) with the parameter values deduced 
by fitting (parameters of the green curves) 
are reasonably small in the whole bias range $\vert V \vert < 1.25\,V_t$ shown (which mimics
the bias range of experimental interest).

\figurename\ref{fig:iv-e0-gamma=0}b depicts the 
energy offset values $\varepsilon_{0}^{fit} = \varepsilon_{0}^{fit}(V_b)$ obtained by 
fitting the $I-V$ curve computed \emph{via} eqn (\ref{eq-i-ib}) up to biases $V_b$ 
indicated on the $x$-axis. As visible there, the fitting parameter $\varepsilon_{0}^{fit}$ 
does significantly depend on the bias range. To get reasonably accurate estimates 
(\emph{i.e.}, $\varepsilon_{0}^{fit} \approx \varepsilon_{0}$), the bias range 
employed for fitting ($V_b$) should be sufficiently narrow 
(as is the case in \figurename\ref{fig:iv-e0-gamma=0_Vmax}). This may seem
unexpected: in principle, a better fit may be expected when more data are sampled. 
In fact, this is surprising only at the first sight; the data to be fitted here are
ideal data resulting from computations via eqn (\ref{eq-i-ib}) that mimic 
(and actually very accurately 
reproduce) measurements (as visible in \figurename\ref{fig:iv-Tan} and \ref{fig:tvs-pvs-Tan}, 
or elsewhere \cite{Baldea:2012a,Baldea:2012b,Baldea:2012h,Baldea:2013b,Baldea:2015d}),
but are not affected by (statistical or measurement \cite{Baldea:2015b}) errors.
The accuracy of the $\varepsilon_{0}^{fit}$-estimates 
obtained by choosing small $V_b$-values as
visible in \figurename\ref{fig:iv-e0-gamma=0}b
is related to the possibility to
accurately ``detect'' slight deviations from linearity in the data to be fitted.
This poses no problem in cases where ``ideal'' data not affected by errors are used
(like those utilized to generate \figurename\ref{fig:iv-e0-gamma=0}b).
However accurate eqn (\ref{eq-i-ib}) is, in contrast to the situation analyzed in 
\figurename\ref{fig:iv-e0-gamma=0}b, real $I-V$ measurements are affected by inherent experimental
errors, and data in a sufficiently broad $V_b$-range are needed for reliable fitting. 
The noise of experimental $I-V$-curves, which is the typical situation 
for STM setups \cite{Guo:11,Wandlowski:11e,Scheer:12}, acts detrimentally 
when too narrow bias ranges are employed for fitting.

\begin{figure}
\centerline{
\includegraphics[width=0.45\textwidth]{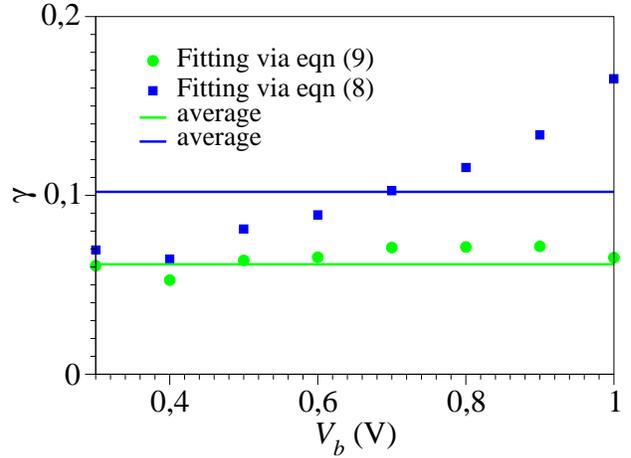}
}
\caption{
The $V_b$ dependence of the 
$\gamma$-values obtained by fitting the experimental $I-V$ curve \cite{Tan:10} 
of \figurename\ref{fig:iv-Tan} \emph{via} eqn (\ref{eq-i3-asym-off})
and (\ref{eq-i-ib}) (blue and green symbols, respectively) within bias ranges $(-V_b, V_b)$,
where $V_b$ is the variable on the abscissa.
Notice the scattering of the blue symbols around average
(much more pronounced than those of the green symbols, which are within 
experimental errors), which reflects the fact that the cubic 
approximation, eqn (\ref{eq-i3-asym-off}), represents an unsatisfactory description:
model parameter values should \emph{not} depend on how broad is the 
bias range employed for fitting.
}
\label{fig:gamma-Tan}
\end{figure}

The unacceptably strong dependence of the fitting parameters on the bias range 
encountered above for cases where $\gamma=0$ becomes even more problematic in the case of
asymmetric curves ($I(-V) \neq -I(V)$, $\gamma \neq 0$). To illustrate this fact, 
in \figurename\ref{fig:gamma-Tan} we depict results showing the dependence on the bias range
($V_b$) of the $\gamma$-parameter obtained by fitting the \emph{experimental} $I-V$ curve 
\cite{Tan:10} with the aid of the ``exact'' eqn (\ref{eq-i-ib}) and the cubic approximation,
eqn (\ref{eq-i3-asym-off}). They are shown as green and blue symbols, respectively.
The difference between the two sets is obvious; while the $V_b$-dependence of 
$\gamma$-values obtained \emph{via} eqn (\ref{eq-i-ib}) is insignificant,  
$\gamma$-values obtained \emph{via} eqn (\ref{eq-i3-asym-off}) vary by a factor $\sim 3$.
It is worth noting in this context that accurate estimates of the parameter $\gamma$
are needed to adequately describe the current asymmetry upon bias polarity reversal 
(``current rectification''). \figurename\ref{fig:iv-Tan} presents a situation where 
the cubic approximation seems reasonable for \emph{one} bias polarity 
(small differences between the green and blue curves for $V<0$) 
but is totally unsatisfactory for the 
other bias polarity ($V>0$). Cases of (inadequate) 
methods able to describe one bias polarity while failing for the opposite bias polarity 
have been presented earlier; see Fig.~5 of ref.~\citenum{Baldea:2012a} and the discussion related to it.

Strong dependencies of the model parameters obtained by fitting using cubic expansions 
similar to eqn (\ref{eq-i3-asym}) have been found earlier in ref.~\citenum{Scheer:12}
and ascribed to a limited applicability of the single-level model. 
Certainly, such limitations cannot be 
ruled out in some cases. However, the present investigation suggests a different possibility:
the single-level description may apply (eqn (\ref{eq-j-arctan}) or (\ref{eq-i-ib})) 
while the related cubic approximations (eqn (\ref{eq-i3-asym}) or (\ref{eq-i3-asym-off}))
fail because are employed for too broad $V$-ranges where terms beyond the third order are important.
\subsection{Bias-driven molecular orbital energy shift. \ib{I. State of the art}}
\label{sec:gamma}
The most common view of current rectification is that the applied bias yields an energy shift of 
the dominant molecular orbital according to eqn (\ref{eq-gamma}). 

To describe this bias-driven shift, a series of studies resort to a simplification 
\cite{CuevasScheer:10,Lee:11,Scheer:12}; 
namely, they relate the asymmetric shift of the molecular orbital energy $\gamma$ to the 
molecule-electrode couplings $\Gamma_{s,t}$
\begin{equation}
\label{eq-gamma-CS}
\gamma = \frac{1}{2}\frac{\Gamma_s - \Gamma_t}{\Gamma_s + \Gamma_t}
\end{equation}
While this procedure reduces the number of fitting parameters, one should be aware that 
this is an \emph{ad hoc} hypothesis without microscopic justification \cite{Lee:11}.
The fact that current rectification is not (necessarily \cite{Baldea:2013d,Baldea:2015a}) 
a result of the asymmetry of molecule-electrode couplings ($\Gamma_s \neq \Gamma_t$) and, 
contrary to what eqn (\ref{eq-gamma-CS}) claims, $\gamma$ should be treated as a model
parameter \emph{independent} of $\Gamma_{s,t}$ has been emphasized earlier in a series 
of works \cite{Datta:04d,Baldea:2012b,Baldea:2013d,Baldea:2014f,Ratner:15a,Baldea:2015a}.

If eqn (\ref{eq-gamma-CS}) applied, the parameter 
\ib{
\begin{equation}
\label{eq-delta}
\delta \equiv \Gamma_{t}/\Gamma_{a}
\end{equation}
}
which quantifies the asymmetry of the molecule-electrode couplings, and the parameter $\gamma$
would depend on each other \cite{different-delta}
\begin{equation}
\label{eq-gamma-delta-CS}
\gamma = \frac{1-\delta}{2}
\end{equation}
Values of the parameters $\gamma$ and 
$\delta$ 
have been estimated by
quantitatively analyzing various experimental data measured under STM platforms 
\cite{Baldea:2013d,Baldea:2014f}; 
the values found there do not satisfy eqn (\ref{eq-gamma-delta-CS}).

For further illustration, we present here another example. 
If eqn (\ref{eq-gamma-CS}) applied, the value 
$\delta = 0.0151$ 
deduced 
in ref.~\citenum{Baldea:2014f} would correspond to $\gamma = 0.49245$. For typical 
low bias conductance values for single-molecule junctions $G/G_0\sim 10^{-3}-10^{-4}$
and biases $e V/\vert \varepsilon_0\vert \approx 1$ 
(typical $\vert \varepsilon_0\vert $-estimates are $\sim 0.5 - 1 $\,eV 
\cite{Wandlowski:11e,Scheer:12,Baldea:2012a,Baldea:2013b,Baldea:2015d}), 
current rectifications of $\sim 15-49$ would result, which is considerably larger 
not only than achieved in the experimental case considered \cite{Baldea:2014f} 
but also in general \cite{Metzger:15}.
 
Another category of works ascribed the bias-driven shift of the energy level
to an asymmetric location of the relevant molecular orbital in the space between 
electrodes. 
\ib{In this picture, the potential $V(z)$ is assumed to drop linearly 
between electrodes. By assuming that the left contact located at $z=-d/2$ 
has the potential $V(-d/2) = + V/2$ and the right contact located at $z=+d/2$  
has the potential $V(+d/2) = - V/2$
(\figurename\ref{fig:homo-c8t}b), 
the potential profile across the junction can be expressed as 
\begin{equation}
\label{eq-linear-drop}
V(z) = - V\frac{z}{d}
\end{equation}
Therefore, to lowest order ($\delta \varepsilon_0 = \mathcal{O}(V)$) the 
energy correction for an MO having its ``center of gravity'' at $z=z_0$ is
} 
\begin{equation}
\label{eq-gamma-lever}
\delta \varepsilon_0 \equiv \varepsilon_{0}(V) - \left . \varepsilon_{0}(V)\right \vert_{V=0} 
= - e V\left(z_0\right) = e V \frac{z_0}{d} \to \gamma = \frac{z_0}{d}
\end{equation}
\ib{
Formulated in words the ``lever rule'' \cite{Metzger:01b} of eqn (\ref{eq-gamma-lever})
expresses the fact that, upon applied bias,  
the MO energy changes according to the change in the local value of the electric potential, which is assumed 
to vary linearly across the junction and is closer to the potential of the closer electrode.
For the case depicted in \figurename\ref{fig:homo-c8t}b, the MO center of gravity 
is closer to the right electrode ($z_0 > 0$); the MO is shifted upwards, following the upward change 
($+eV/2$) in the Fermi energy of the right electrode, by an amount determined by the 
MO fractional position $z_0/d$. 
} 
\subsection{\ib{Bias-driven molecular orbital energy shift. II.
A counterintuitive example backed by quantum chemical calculations}}
\label{sec:qc}
\begin{figure}
$ $\\[-12ex]
\centerline{\includegraphics[width=0.35\textwidth]{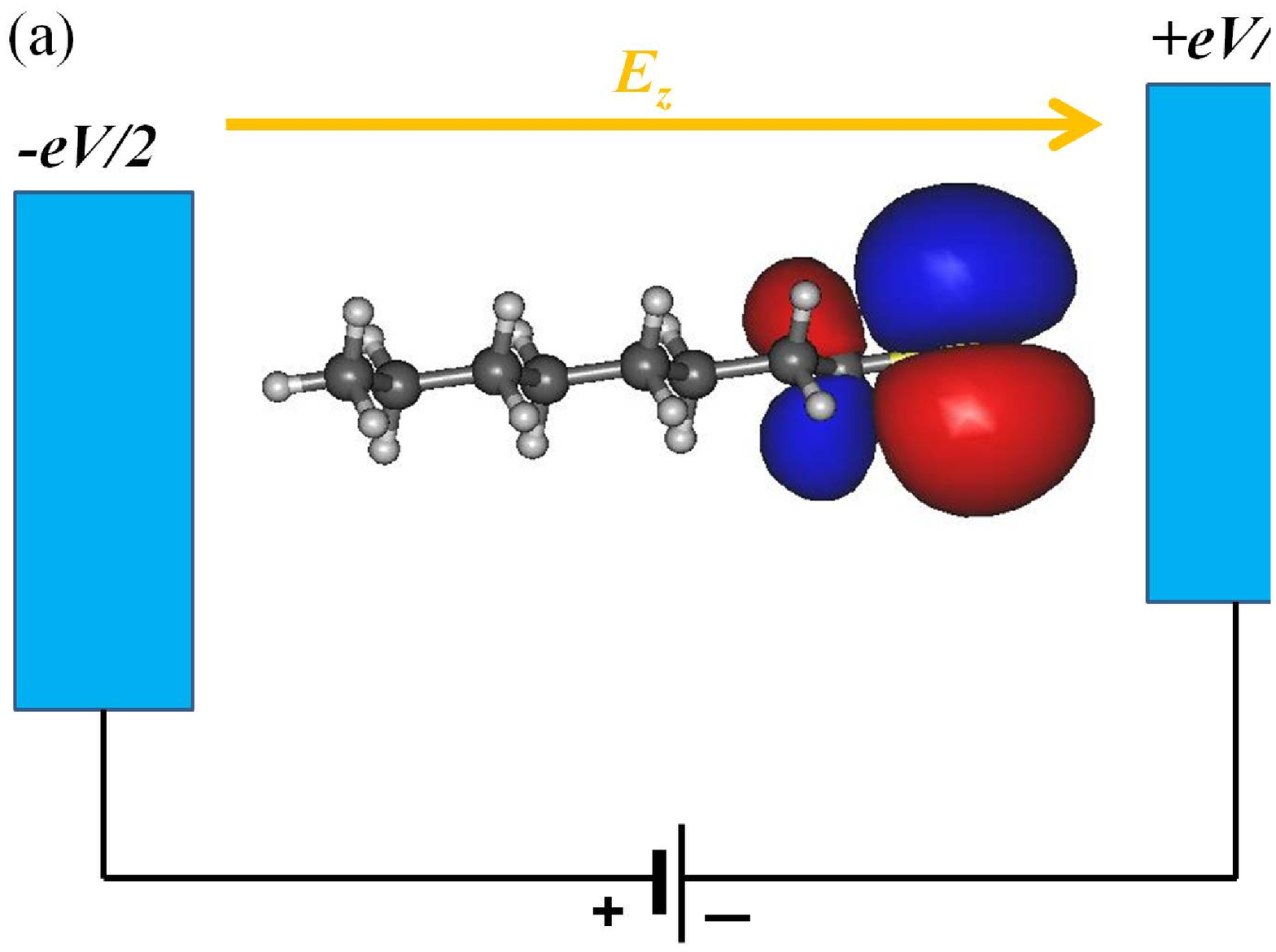}}
$ $ \\[-26ex]
\centerline{\includegraphics[width=0.35\textwidth]{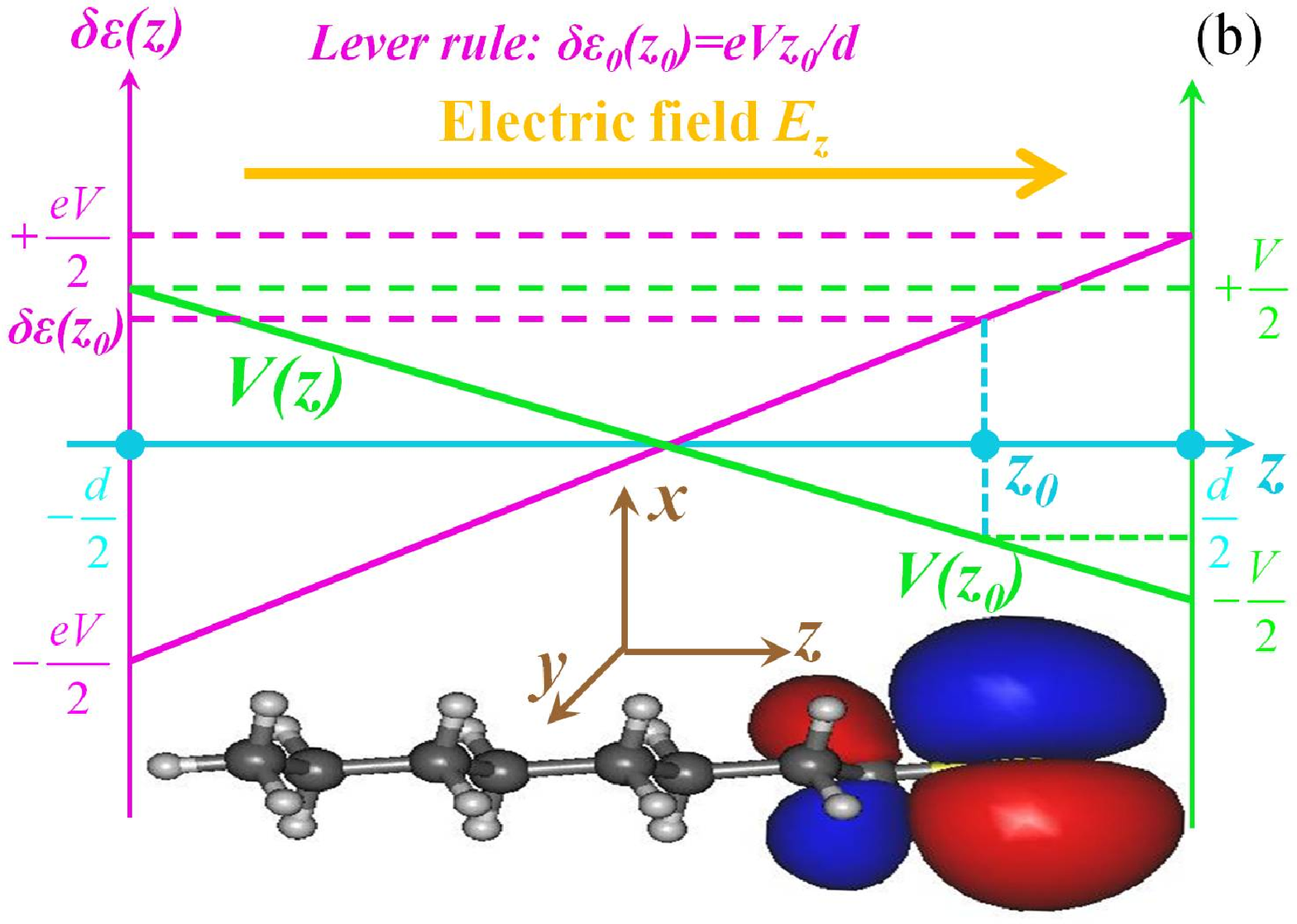}}
$ $\\[-9ex]
\centerline{\includegraphics[width=0.35\textwidth]{fig6c_revisited}}
$ $\\[-5ex]
\caption{(a) Schematic representation of an octanethiol molecule \ce{CH3(CH2)7SH} in external electric 
field $E_z$.
(b) The spatial distribution of the highest occupied molecular orbital (HOMO) 
and the energy shift $\delta\varepsilon_{0}\left(z_0\right)$ 
(upwards in this figure) expected within the ``lever-rule'' argument based on the assumption 
of a linear drop of the potential $V(z)$ (green solid line) across the electrodes.
(c) The HOMO energy in an applied electric field $E_z$ computed within various methods specified 
in the legend: OVGF, second (2P)- and third (3P)-order pole approximation, Hartree-Fock (HF)
and Kohn-Sham (KS) HOMO energies. Notice that although the HOMO density is concentrated near the 
thiol group \ce{SH}, it is the more distant electrode that prevails in shifting the HOMO energy. 
}
\label{fig:homo-c8t}
\end{figure}

It is worth emphasizing that the ``lever rule'' (schematically 
depicted in \figurename\ref{fig:homo-c8t}b), which
justifies the term of potential profile asymmetry or voltage division factor 
\cite{Datta:03} used for the parameter $\gamma$, assumes a linear potential drop across
the junction. In some cases, data could be quantitatively analyzed within this picture
validated by inserting in a controlled way spacers in the molecules 
embedded in junctions \cite{Whitesides:10a}. 

However, by assuming a linear potential profile, screening effects are neglected.
To demonstrate that a negligible screening is a fact that can by no means
be taken for granted, we present below (\figurename\ref{fig:homo-c8t}) 
results for the HOMO energy of 
the alkanethiol molecule \ce{CH3(CH2)7SH} placed in an external field $E_z$ along the 
($z$-)molecular axis. 

\ib{These results have been obtained \emph{via} genuine \emph{ab initio} quantum chemical calculations
(OVGF and CCSD, \emph{vide infra}).
Our aim is to bring surprising aspects to experimentalists' attention 
when they process molecular transport data.
Therefore, to make this subsection accessible to a broader audience 
some relevant details will be given below in order to justify why such 
\emph{ab initio} quantum chemical calculations
beyond the widely employed density functional theory (DFT) are needed and how they are performed.

DFT calculations are very useful to obtain a variety of ground state properties. For geometry 
optimization, such calculations based on the 
B3LYP functional as implemented in GAUSSIAN 09 \cite{g09} have also been performed in this paper.
However, 
}
as well documented \cite{Parr:89,Kohn:96}, the Kohn-Sham (KS) ``orbitals'' utilized in DFT calculations 
are \emph{not} physical molecular orbitals. Less problematic conceptually is the HOMO.
\ib{\emph{If} one knew the \emph{exact} exchange-correlation functional (the key DFT quantity), the KS-HOMO energy 
would correspond to the lowest ionization energy \cite{Almbladh:85}. However,}
for typical molecules used to fabricate
molecular junctions \ib{this is not the case}; 
even the HOMO energy is poorly described within the DFT \cite{Baldea:2014c}.

To avoid this issue, here we present results for the HOMO energies 
obtained \emph{via} genuine \emph{ab initio} 
quantum chemical 
calculations based on the outer valence Green's function (OVGF) method \cite{Cederbaum:77,Schirmer:84}.
\ib{The OVGF method is a diagrammatic many-body approach \cite{fetter}, wherein the self-energy entering 
the electronic Dyson equation includes full (i) second- and (ii) 
third-order terms of electron-electron interaction. Moreover, (iii) it is augmented 
by a geometrical approximation (physically associated to a screening factor) to also partially 
include fourth- and higher-order corrections \cite{sub-series}.
The ionization energies are determined from the poles of the Green's function computed in this way.
In \figurename\ref{fig:homo-c8t}c the labels 2P, 3P, and OVGF refer to the lowest ionization energy
with reversed sign (HOMO energy) corresponding to the methods denoted above by 
(i), (ii), and (iii), respectively.
} 

\ib{To obtain the dependence on the electric field $E_z$ applied along the molecule 
of the HOMO energy (\figurename\ref{fig:homo-c8t}c), 
calculations at the OVGF/6-311++g(d,p) level of theory have been performed.} 
Such quantum chemical calculations are known to be accurate 
not only for medium size molecules like the presently considered alkanethiol molecule 
but also for larger molecules (like \ce{C60} \cite{Ortiz:14}).
As expected \cite{Baldea:2014c}, differences between the OVGF HOMO energies and the 
Kohn-Sham ``energies'' (also shown in \figurename\ref{fig:homo-c8t}c) are very large. 
Differences between the OVGF energies and the Hartree-Fock (HF) values and
those obtained within the second-order pole approximation (2P) are also 
significant, while the third-order (3P) pole approximation appears to be accurate 
in this case. 

We applied the OVGF method because this is the most accurate approach
to compute ionization (and electron attachment relevant for LUMO-mediated conduction) 
energies in the presence of an
external electric field implemented in existing (or, at least, to our disposal) 
quantum chemical packages. 
However, the example presented below suggests that there is no 
practical need to resort to even more elaborate
many-body approaches to compute the lowest ionization energy 
(which would be the HOMO
energy with reversed sign \emph{if} the one-particle description applied).

\ib{CCSD (coupled-cluster (CC) singles (S) and doubles (D)) \cite{Stanton:94,Schirmer:09} 
is such an approach; it represents the state-of-the-art of quantum chemistry to 
treat many-electron systems of medium size molecules.  
The CC technique constructs multi-electron wave functions by applying exponential cluster operators on the  
the Hartree-Fock (HF) (molecular orbital) wave function. 
In the specific case of CCSD, the cluster operator is truncated to single 
and double excitations (in short, ``singles'' and ``doubles'').}
Within the CCSD framework, the lowest ionization energy 
can be computed either by applying the equation of motion method (EOM-IP-CCSD) \cite{Stanton:94} or by 
subtracting the total CCSD ground state energies of the cationic and neutral molecular species ($\Delta$-CCSD 
\cite{Baldea:2014c}). As a further check of the OVGF approach, we mention that 
the OVGF-value (9.089\,eV) agrees well agreement with the 
values thus obtained (8.996\,eV using EOM-IP-CCSD and 8.945\,eV using $\Delta$-CCSD) without applied
electric field. 

CCSD calculations of this work have been performed with CFOUR \cite{cfour},
a package also utilized to compute the spatial distribution of the HOMO 
(see ref.~\citenum{Baldea:2014c} for details). 
Again, to avoid issues related to KS orbitals, we have calculated the natural orbital expansion
of the reduced density matrix at the EOM-IP-CCSD level, as the most reliable approach to characterize
the spatial distribution of the extra hole (or electron in cases of LUMO-mediated conduction).
By inspecting the natural orbital expansion, we found that the extra hole is almost entirely 
($\sim 97\%$) concentrated in a single natural orbital (``HOMO''). It is this 
spatial distribution that is shown in \figurename\ref{fig:homo-c8t}a and b.

The most important finding of this subsection emerges from the comparison of
panels b and c of \figurename\ref{fig:homo-c8t}:
the OVGF method (as well as the other methods related to it discussed above) predicts 
HOMO energies clearly exhibiting a trend opposite to the ``lever rule'' expectation.
In view of this fact, namely, that cases exist, where the ``lever rule'' may fail, 
rather than voltage division factor or potential profile asymmetry,
Stark effect strength \cite{Xue:03} 
may be a possible, more appropriate term when referring to the parameter $\gamma$. 

\ib{
Although a linear dependence of $\varepsilon_0$ on the applied bias is
often assumed in transport studies (also in electrochemical context 
\cite{Alessandrini:06,Baldea:2013d}), we are not aware of 
any quantum chemical study reporting such a result for molecules often used to fabricate 
molecular junctions. Therefore, we believe that the strict linearity of the dependence 
of the HOMO energy on the applied field/bias represents an important result of the present paper. 
Noteworthy, the results of the OVGF-computations 
(depicted by points in \figurename\ref{fig:homo-c8t}c), which perfectly lie on a straight line,
correspond to electric field values up to 2\,V/nm. These values safely cover
the typical experimental range for molecular devices, which is in most cases up to about 
1\,V/nm, since beyond this value field ionization may become significant.} 
\section{Conclusion}
\label{sec:conclusion}
An important finding of the present paper is the demonstration that current transport data 
processing based on cubic expansions of the current as a function of voltage
is inappropriate. First, this
typically underestimates the energy offset of the dominant molecular orbital
by a factor of about two. Because DFT calculations 
typically underestimate the HOMO-LUMO gap by a similar factor, in the light of the present finding, 
``agreements'' between experiments
and theories using Kohn-Sham orbital energies \emph{uncorrected} by employing more accurate quantum 
chemical methods and/or image charge effects could/should be reconsidered. 
Second, the application of the cubic expansion for bias ranges of experimental interest (almost inherently)
yields parameter values depending on how broad is the bias range employed for fitting;
this may easily be interpreted as an unphysical result, 
creating the impression that the single level description is invalid. In reality, more plausible
is that the cubic expansion rather than the single-level description is inadequate.
A third drawback of the cubic expansion is its inability to quantitatively describe \emph{asymmetric} 
$I(V) \neq - I(-V)$ curves. This is revealed by the fact that, contrary to experiments \cite{Tan:10},
it yields transition (peak) voltages of equal magnitude for both bias polarities 
(\emph{cf.}~\figurename\ref{fig:tvs-pvs-Tan}); this is an aspect on which 
a theoretical simulation presented in ref.~\citenum{Baldea:2015c} already drew attention. 

Another important finding reported in this paper is the fact that the 
bias-driven shifts of molecular orbital energies are necessarily determined neither by the asymmetry 
of the molecule-electrode couplings nor to the asymmetric location of the ``center of gravity'' of the
molecular orbitals relative to the two electrodes. 
The latter aspect is important also because it emphasizes that even if 
a single orbital dominates the charge transport through a certain molecular junction, other molecular
orbitals can have indirect contributions \emph{via} subtle screening effects that may yield counterintuitive
effects of the kind presented above.
\ib{We chose to present a single (counter-)example, namely, the case of an isolated
benchmark molecule (octanethiol) in external field. We could present more (counter-)examples (\emph{e.g.},
a fuller class of alkanethiols), but this would not add any further evidence; 
we do by no means claim that the ``lever rule'' fails in \emph{all} cases. 
To avoid ambiguity, we considered an \emph{isolated} molecule.
If we presented a molecule linked to electrodes, never ending questions might arise, \emph{e.g.} on the  
contacts' geometry (atop, bridge, hollow) or nature (chemisorption \emph{vs.}~physisorption).
What is important for the present purpose is to show that the (upward or downward) MO shift due to an applied
electric field is not necessarily directly related to the MO location.}
\section*{Acknowledgment}
The author thanks Pramod Reddy for providing him the raw data utilized in 
\figurename\ref{fig:iv-Tan} and \ref{fig:tvs-pvs-Tan}.
Financial support provided by the Deu\-tsche For\-schungs\-ge\-mein\-schaft 
(grant BA 1799/2-1) is gratefully acknowledged.
\renewcommand\refname{Notes and references}
\footnotesize{
\bibliographystyle{rsc}
\providecommand*{\mcitethebibliography}{\thebibliography}
\csname @ifundefined\endcsname{endmcitethebibliography}
{\let\endmcitethebibliography\endthebibliography}{}

}
\end{document}